\documentclass[12pt]{article}

\usepackage{amsmath,amssymb}
\usepackage[dvips]{graphicx,color}
\usepackage[round]{natbib}
\usepackage{mathrsfs}
\usepackage{bm, xcolor}
\usepackage{mathrsfs}
\usepackage{verbatim}
\usepackage{threeparttable}
\usepackage{color}
\usepackage[colorlinks,
linkcolor=blue,
anchorcolor=blue,
citecolor=blue]{hyperref}
\usepackage[utf8]{inputenc}
\usepackage{epstopdf}
\usepackage{booktabs}
\usepackage{graphicx}
\usepackage{subfigure}
\usepackage{float}
\usepackage{listings}
\usepackage{amssymb}
\usepackage{amsthm}
\usepackage{multirow}
\usepackage{rotating}
\usepackage{longtable}
\usepackage{enumitem}
\usepackage{threeparttablex} 
\usepackage{booktabs} 
\usepackage{amssymb}
\usepackage{algorithmic}
\usepackage{algorithm}
\usepackage{cases}

\usepackage{caption}
\usepackage{filecontents}
\usepackage[section]{placeins}

\makeatletter
\def\singlespace{\def\baselinestretch{1}\@normalsize}

\@addtoreset{equation}{section}

\renewcommand{\hat}{\widehat}

\def\singlespace{\def\baselinestretch{1}\@normalsize}

\makeatother

\def\E{\mbox{E}}

\def\D{{\mathcal D}}

\newdimen\biblioindent    \biblioindent=30pt

 at 7truept

\newtheorem{theorem}{Theorem}[section]
\newtheorem{corollary}{Corollary}[section]

\newtheorem{proposition}{Proposition}[section]

\newtheorem{remark}{Remark}[section]

\def\boxit#1{\vbox{\hrule\hbox{\vrule\kern6pt\vbox{\kern6pt#1\kern6pt}\kern6pt\vrule}\hrule}}

\numberwithin{equation}{section}
\usepackage{chngcntr}

\usepackage{xr}

\begin{document}
	
	\renewcommand{\baselinestretch}{1.3}

	\title {\bf Test and Measure for Partial Mean Dependence Based on Machine Learning Methods}
	\author{Leheng Cai$^1$, Xu Guo$^2$ and Wei Zhong$^3$
	\thanks{We would like to thank the Editor, the Associate Editor, and the  anonymous reviewers for their valuable comments and constructive suggestions, which lead to significant improvements in the paper. All authors equally contributed to this article, and the authors are listed in the alphabetical order. Corresponding authors: Xu Guo, Wei Zhong. Email address: xustat12@bnu.edu.cn; wzhong@xmu.edu.cn.}\\
 {\small \it $^{1}$ Tsinghua University, $^{2}$ Beijing Normal University and $^{3}$ Xiamen University}\\
	}
	\date{}
	
	\date{}
	
	\maketitle
	\renewcommand{\baselinestretch}{1.5}
	
\begin{abstract}

\medskip
		
It is of importance to investigate the significance of a subset of covariates $W$ for the response $Y$ given covariates $Z$ in regression modeling. To this end, we propose a significance test for the partial mean independence problem based on machine learning methods and data splitting. The test statistic converges to the standard chi-squared distribution under the null hypothesis while it converges to a normal distribution under the fixed alternative hypothesis. Power enhancement and algorithm stability are also discussed. If the null hypothesis is rejected, we propose a partial Generalized Measure of Correlation (pGMC) to measure the partial mean dependence of $Y$ given $W$ after controlling for the nonlinear effect of $Z$. We present the appealing theoretical properties of the pGMC and establish the asymptotic normality of its estimator with the optimal root-$N$ convergence rate. Furthermore, the valid confidence interval for the pGMC is also derived. As an important special case when there are no conditional covariates $Z$, we introduce a new test of overall significance of covariates for the response in a model-free setting. Numerical studies and real data analysis are also conducted to compare with existing approaches and to demonstrate the validity and flexibility of our proposed procedures. 
		
		
\end{abstract}

	\baselineskip=24pt
	
	\pagestyle{plain}
	
	\vspace{3mm}
	
\section{Introduction}
It is of importance to investigate the significance of a subset of covariates for the response in regression modeling. In practice, some covariates are often known to be related to the response of interest based on historical analysis or domain knowledge. We aim to test the significance of another group of covariates for the response given the known covariates, especially in high dimensional data. For example, \cite{2006Regulation} has detected 22 gene probes that are related to human eye disease from 18,976 different gene probes and it is interesting to test whether the remaining high dimensional gene probes still contribute to the response conditional on the subset of 22 identified gene probes. This problem can be formulated as the following hypothesis testing problem
\begin{equation}\label{partialmeantest}
	H_0:\,\,\,\E(Y|X)=\E(Y|Z), \mbox{~v.s.~~~} H_1:\,\,\,\E(Y|X)\ne \E(Y|Z),
\end{equation}
where $Y$ is a scalar-valued response variable of interest, $X=\left(X_1,\cdots,X_p\right)^\top=\left(Z^\top,W^\top\right)^\top\in\mathbb{R}^p$ is a $p-$dimensional vector of covariates, $Z\in\mathbb{R}^{p_1}$ and $W\in\mathbb{R}^{p_2}$ are two subsets of $X$ and $p_1+p_2=p$.
The equality in $H_0$ indicates that $Y$ and $W$ are partially mean independent after controlling for the nonlinear effect of $Z$. If $H_0$  is not rejected, $W$ is considered as insignificant for  $Y$ and can be omitted from the regression model. Thus, this test is also called a significance test \citep{2001significance} or an omitted variable test \citep{1996consistent}. 
In the literature, several tests in nonparametric settings have been proposed based on locally and globally smoothing methodologies whose representatives are \cite{1996consistent}'s test and \cite{2001significance}'s test, respectively. However, both tests are based on kernel smoothing estimation and severely suffer from the curse of dimensionality.
\cite{zhu2018dimension}  proposed a dimension reduction-based adaptive-to-model (DREAM) test for the significance of a subset of covariates.
However, the DREAM test only focuses on the relatively low dimensional situation and cannot perform satisfactorily in high dimensions. Therefore, it is still challenging to test $H_0$  in high dimensional settings.


In this paper, we propose a new significance test for the partial mean independence problem in  (\ref{partialmeantest}) based on machine learning methods. 
The obstacle of testing $H_0$ is estimating the conditional means in high dimensions which makes the classical nonparametric estimation such as kernel smoothing and local polynomial fitting perform poorly. {Alternatively, we apply the machine learning methods such as deep neural networks (DNN) and  eXtreme Gradient Boosting (XGBoost) to estimating the conditional means in the new test.} Specifically, we randomly split the data into two parts. We estimate the conditional means via machine learning methods using the first part of the data and conduct the test based on the second part. We theoretically show that the test statistic converges to the standard chi-squared distribution under the null hypothesis while it converges to a normal distribution {under the fixed alternative hypothesis.
In this paper, we introduce a new reformulation of the null hypothesis, which is helpful to obtain a tractable limiting null distribution. Power enhancement and algorithm stability based on multiple data splitting are also discussed. }

Recently, there is an increasing interest in developing powerful testing procedures for the partial mean independence problem. Some remarkable developments include \cite{dai2022significance}, \cite{williamson2023general}, \cite{verdinelli2021decorrelated}, and \cite{lundborg2022projected}. These papers also adopt machine learning methods and sample splitting. To deal with the fundamental difficulty that a naive test statistic has a degenerate distribution, different attempts are made. \cite{dai2022significance} added random noise to their test statistic. \cite{williamson2023general} alleviated this problem by estimating two non-vanishing parameters on separate splits of the data which was also adopted in the two-split test of \cite{dai2022significance}. The method of \cite{verdinelli2021decorrelated} is similar to  \cite{dai2022significance} but without introducing additional randomness. They achieved tractable limiting null distributions at the price of possible power loss. \cite{lundborg2022projected} introduced a notable procedure named Projected Covariance Measure and proved that under linear regression models with fixed dimensional $W$, their procedures are more powerful than those of \cite{dai2022significance} and \cite{williamson2023general}. However, under the model-free framework, which is the focus of this paper, their theory requires three or more subsamples. For other relevant papers, see also \cite{lei2018distribution}, \cite{zhang2020floodgate}, \cite{shah2020hardness}, \cite{gan2022inference}, \cite{tansey2022holdout}, and \cite{verdinelli2023feature}. For a recent review, see \cite{lundborg2023modern}.

In this paper, we introduce a new procedure to deal with the degenerate null distribution issue.
{Our introduced test procedure has the following four merits compared with other existing methods. Firstly, the newly proposed test statistic does not involve data perturbation as in \cite{dai2022significance} nor require three or more subsamples as in \cite{williamson2023general}. Secondly, the proposed procedure has nontrivial power against the local alternative hypothesis which converges to the null faster than those in \cite{williamson2023general} and \cite{dai2022significance}. Thirdly, our test procedure only requires estimating two conditional means, and thus it is computationally practical, while Algorithm 1 in \cite{lundborg2022projected} necessitates conducting estimations of at least five conditional means. Furthermore, we novelly introduce a power enhancement procedure, which significantly differs from the power enhancement technique using $L_{\infty}$ type thresholding statistic in \cite{fan2015power}, turning the degeneracy issue into a blessing of power.}

\par If the null hypothesis $H_0$ is rejected,  then it is meaningful to measure the partial mean dependence of $Y$ given $W$ after controlling for the nonlinear effect of $Z$. The traditional Pearson's partial correlation coefficient  cannot measure the nonlinear partial dependence between $Y$ and $W$ given $Z$. In the literature, several works focus on quantifying the partial/conditional dependence between two random objects given another one. \cite{2014partialdc} defined the partial distance correlation  with the help of the Hilbert space for measuring the partial dependence. \cite{wang2015conditional} introduced a conditional distance correlation for conditional dependence. \cite{azadkia2021simple} proposed a new conditional dependence measure which is a generalization of a univariate nonparametric measure of regression dependence in \cite{2013dette}.
However, although conditional independence implies conditional mean independence, the inverse is not right and the conditional mean independence has a clear connection to regression modelling. To measure the partial mean dependence, \cite{2015partial} extended the idea of \cite{2014partialdc} to define the partial Martingale Difference  Correlation (pMDC) to measure the conditional mean dependence of $Y$ given $W$ adjusting for $Z$. However, asymptotic properties of the pMDC are still unclear. As another main contribution, we propose a new partial dependence measure, called partial Generalized Measure of Correlation (pGMC), and derive its theoretical properties and construct confidence interval. The new pGMC is based on the decomposition formula of the conditional variance and thus can be considered as an extension of Generalized Measure of Correlation (GMC) proposed by \cite{zheng2012generalized}. The new pGMC has several appealing properties. First, it takes value between 0 and 1. The value is 0 if and only if $Y$ and $W$ are conditionally mean independent given $Z$, and is 1 if and only if $Y$ is equal to a measurable function of $W$ given $Z$. Second, the asymptotic normality  of the proposed estimator of the pGMC is derived with the optimal convergence rate and the associated confidence interval is also constructed. 

As an important special case when there is no  conditioning random object $Z$,   (\ref{partialmeantest}) becomes
\begin{equation}\label{meantest}
	\widetilde{H}_0:\,\,\,\E(Y|X)=\E(Y), \mbox{~v.s.~~~} \widetilde{H}_1:\,\,\,\E(Y|X)\ne \E(Y),
\end{equation}
where the null hypothesis $\widetilde{H}_0$  means the conditional mean of $Y$ given $X$ does not depend on $X$. It is  a fundamental testing problem to check the overall significance of covariates $X$ for modeling the mean of the response $Y$ in regression problems. For example, the F-test is the classical and standard procedure to determine the overall significance of predictors in linear models. Without parametric model assumptions, many model checking procedures can also be applied, for instance, \cite{wang2006testing}, \cite{gonzalez2013updated}, and \cite{guo2016model}. In high dimensions, many testing procedures including \cite{goeman2006testing}, \cite{zhong2011tests}, \cite{guo2016tests}, and \cite{cui2018test} have been developed to assert the overall significance for linear models or generalized linear models.
To accommodate the high dimensionality in the model-free setting, both \cite{zhang2018conditional} and \cite{li2023testing} considered a weak but necessary null hypothesis
\begin{equation}\label{weaknull}
	\widetilde{H}_0^\prime:\,\,\,\E(Y|X_j)=\E(Y), \mbox{ almost surely, for all } 1\leq j \leq p,
\end{equation}
and proposed the sum-type test statistics over all marginal tests.
In this paper, we also propose a new test of overall significance for $\widetilde{H}_0$. 

The paper is organized as follows. In section 2, we propose
a new significance test for the partial mean independence. In section 3,
a new partial Generalized Measure of Correlation (pGMC) is introduced. Section 4 considers the important special case  to test the overall significance for $\widetilde{H}_0$. 
Numerical studies are conducted in section 5. In section 6, we present two real data examples. 
Additional simulation studies and the proofs of theoretical results are presented in the supplementary materials.

\section{Partial Mean Independence Test (pMIT)}\label{sec2}
\subsection{Test statistic and its asymptotic distributions}
In this section, we consider the partial mean independence test problem (\ref{partialmeantest}). That is, it aims to check whether $Y$ and $W$ are partially mean independent after controlling for the nonlinear effect of $Z$. {We first make a new reformulation of the null hypothesis $H_0$ as follows
\begin{align}\label{partialmeantest2}
	{H}_0^\prime:\,\,\,\E(\tilde Y|X)=\E(\tilde Y),
\end{align}
where $\tilde Y=Y-\E(Y|Z)$.}
{Note that $\E(\tilde Y)=\E[Y-\E(Y|Z)]=0$, and thus the right-hand side of $H'_0$ is actually reduced to 0. Based on this reformulation $H'_0$, a natural idea is to compare the estimators of $\E(\tilde Y|X)$ and $\E(\tilde Y)$ via measuring whether $\E\left[\{\E(\tilde Y|X) - \E(\tilde Y)\}^2\right]$ equals 0. By introducing the sample analog estimator of the zero term $\E(\tilde Y)$,  one could avoid the fundamental difficulty that a naive test statistic has a degenerate distribution.} 
 Let $\D=\{X_i, Y_i\}_{i=1}^{N}$ be a random sample from the population $\{X,Y\}$. We propose a new test statistic based on machine learning methods and data splitting technique. To be specific, we split the data randomly into two independent parts, denoted by $\D_1$ and $\D_2$. Let $|\D_2|=n_2=:n$ and $|\D_1|=n_1=:n^\gamma$ with $\gamma>1$, where $n_1+n_2=n^\gamma+n=N$. Without loss of generality, we assume that $n^{\gamma}$ is an integer. 
{We first estimate the conditional mean  $h(Z_i)=\E(Y_i|Z_i)$  using machine learning methods  based on the first subset of data $\D_1$, denoted by $\hat h_{\D_1}(Z_i)$, and then estimate $g(X_i)=\E(\tilde{Y}_i|X_i)$ by regressing $Y_i-\hat h_{\D_1}(Z_i)$ on $X_i$ using machine learning methods based on $\D_1$ to get its final estimator $\hat g_{\D_1}(X_i)$.} Then, we construct the test statistic via comparing the estimators of $\E(\tilde Y|X)$ and $\E(\tilde Y)$ based on $\D_2$,
\begin{equation} \label{pmittest}
	T_n = \frac{1}{n}\sum_{i\in \D_2}\left[ \hat g_{\D_1}(X_i)-\frac{1}{n}\sum_{j\in \D_2}\left\{Y_j-\hat h_{\D_1}(Z_j)\right\} \right]^2.
\end{equation}
{To simplify notations, here we use $i\in\D_2$ to denote that $i$ belongs to the index set corresponding to $\D_2$.} Intuitively, the larger $T_n$ is, the stronger evidence we have to reject $H_0$. We call this test the partial mean independence test (pMIT).

\begin{remark}
First, the data splitting is crucial to control the type-I error in the pMIT procedure.
Second, the unbalanced sample splitting with a larger training set for the conditional means is necessary to achieve a tractable limiting null distribution for the significance test (\ref{partialmeantest}).  This observation makes the proposed pMIT test distinguished from \cite{cai2022model} which proposed a similar test for (\ref{partialmeantest}) via data splitting. However, \cite{cai2022model} didn't  theoretically investigate either asymptotic distributions nor the sample splitting ratio.
\end{remark}

{To proceed, we make some discussions on the reformulation (\ref{partialmeantest2}).
Different from existing approaches to compare the two conditional expectations $\E(Y|X)$ and $\E(Y|Z)$,  we compare one conditional expectation $\E(\tilde Y|X)$ and one unconditional expectation $\E(\tilde Y)$ of transformed $\tilde Y$.
With the unbalanced sample splitting strategy, under the null hypothesis, the estimation errors from $\sum_{i\in\D_2}\hat g_{\D_1}^2(X_i)$ are controlled and $\sum_{i\in\D_2}\hat g_{\D_1}^2(X_i)$ becomes negligible. While with the formulation $H'_0$, $\sum_{i\in\D_2}(n^{-1}\sum_{j\in\D_2}\tilde Y_j)^2$ converges to a chi-squared distribution and determines the asymptotic distribution of $T_n$, making it valid for inference. If instead, we do not estimate $\E(\tilde Y)$ and consider $n^{-1}\sum_{i\in \D_2}\hat g^2_{\D_1}(X_i)$ directly, this natural statistic still has the degeneracy issue. Thus, introducing the estimation of ${\rm E}(\tilde Y)$ is crucial and is a new complement to existing methods.
Essentially, we tackle the issue of degeneracy by introducing additional randomness in the error term of $Y$. However there are two differences between our procedure and \cite{dai2022significance}'s procedure. Firstly, our procedure does not involve data perturbation and does not require to select perturbation size. Secondly and more subtlely, we introduce the randomness in $\tilde Y$ directly in the comparison with $g(X)$. While \cite{dai2022significance}'s procedure involves data perturbation with the difference of prediction errors of $m(X)$ and $h(Z)$. This seemingly minor change has a major beneficial effect: theoretically our procedure can detect local alternative hypothesis which converges to the null at faster rate. Intuitively, our procedure directly detects shift from $g(X)$, while their procedure focuses on $\E[g^2(X)]$.} Furthermore, our reformulation is general and can be adopted to deal with other conditional mean restriction testing problems. For instance, we may extend our procedure to test the conditional
independence. Actually the conditional independence of $Y$ and $W$ given $Z$ is equivalent to $\E[I(Y\leq y)|X]=\E[I(Y\leq y)|Z]$ for any $y$. We can then apply our method for each $y$ and integrate all the information at different $y$'s to construct the final test statistic.

Then, we derive the asymptotic distributions of the statistic $T_n$. Let $\sigma^2_{Y|Z}=\mbox{Var}\left\{Y-\E(Y|Z)\right\}=\E\left[\left\{Y-h(Z)\right\}^2\right]$ and it can be estimated by
$\hat\sigma^2_{Y|Z}={n}^{-1}\sum_{i\in \D_2}\left\{Y_i-\hat h_{\D_1}(Z_i)\right\}^2.$
We impose the following assumption on the estimation errors of the conditional means using machine learning methods. For some $a\in(0,1)$ and a  generic positive constant $c$, we suppose
\begin{itemize}
	\item[(C1)] $\E\left[\left\{\hat h_{\D_1}(Z)-h(Z)\right\}^2\right]\leq c n_1^{-a}=c n^{-\gamma a}$ and $\E\left[\left\{\hat g_{\D_1}(X)-g(X)\right\}^2\right]\leq c n_1^{-a}=c n^{-\gamma a}$.
\end{itemize}

{The first part of (C1) reasonably assumes that  machine learning methods are able to estimate the conditional mean accurately with a certain convergence rate of $n_1^{-a}$.  For example, \cite{bauer2019deep},
\cite{schmidt2020nonparametric}  and \cite{kohler2021rate} studied the high-level estimation errors of estimators of conditional means via DNNs. Specifically, 
\cite{bauer2019deep} showed that ${\rm E}\left[\left\{\hat h_{\D_1}(Z)-h(Z)\right\}^2\right]=O((\log n_1)^3n_1^{-2q/(2q+d^*)})$, where $q$ is the smoothness parameter and $d^*$ is the intrinsic dimensionality. From their corollary 1, the above rate can be even in the order of $(\log n_1)^3n_1^{-2/3}$. 
{Further, rates of convergence of various boosting procedures have been widely studied. For instance, \cite{buhlmann2003boosting} derived that the estimation error achieves rate of convergence $O(n^{-2q/(2q+1)})$ for one-dimensional regression, when adopting smoothing splines as weak learners. \cite{Buhlmann2006} established the consistency of $L_2$-Boosting in high-dimensional setting without stating the rate. \cite{KUECK2023714} proved that for high-dimensional linear regression model, iterated post-$L_2$-Boosting and orthogonal $L_2$-Boosting achieve the rate of convergence $O(s\log p\log n/n)$ with $s$ being the sparsity level. } More discussions can be seen after Condition (C2). For the second part of (C1), we note that  $g(X_i)={\rm E}(Y_i|X_i)-{\rm E}(Y_i|Z_i)$ can be estimated by $\hat g_{\D_1}(X_i)=\hat m_{\D_1}(X_i)-\hat h_{\D_1}(Z_i)$, where $\hat m_{\D_1}(X_i)$ is an estimator of the conditional mean $m(X_i)={\rm E}(Y_i|X_i)$. Thus, the second part of (C1) is deduced from the first assumption in (C1) and a similar assumption on $\hat m_{\D_1}(X_i)$. 
In the literature, \cite{williamson2021nonparametric} 
assumed that $\E\left[\left\{\widetilde h_{\D_1}(Z)-h(Z)\right\}^2\right]\leq c n_1^{-a}$ with $1/2<a$, where $\widetilde h(Z)$ is obtained by regressing $\hat m_{\D_1}(X)$ on $Z$ based on $\D_1$. \cite{lundborg2022projected} also imposed similar conditions about estimation errors for the relevant estimators in their assumptions 3-4. In a special case, for linear regression model with ordinary least squared estimator, it can be shown that our proposed estimator satisfies $\E\left[\left\{\hat g_{\D_1}(X)-g(X)\right\}^2\right]\leq c n_1^{-1}$. {Further for linear regression model with LASSO estimator, we show that our proposed estimator satisfies $\E[\left\{\hat g_{\D_1}(X)-g(X)\right\}^2]\leq c s\log p/n_1$ with $s$ being the sparsity level. Thus in both cases, the second assumption in (C1) holds.}

{Although we are not constrained to any particular estimation method to estimate $g(X)$, we have found that the proposed one works well in practice. In fact under the null hypothesis, it generally has very small overall estimation error $\sum_{i\in\mathcal D_2}\hat g^2_{\D_1}(X_i)$ and thus can control empirical sizes very well. While under the alternative hypothesis, as long as the over-fitting is not too severe, our procedure can still detect corresponding alternative hypothesis. As in \cite{lundborg2022projected}, we do not claim that our choice is always the best, but we find this choice works well in practice and set it to be a sensible default choice.}

Next, we derive the asymptotic distributions of $T_n$ in the following theorem.
\begin{theorem}\label{theorem1}
	{Assume that $c_Y<{\rm{E}}\left[\left\{Y-h(Z)\right\}^2\right]<C_Y$ for some positive constants $c_Y$ and $C_Y$}, and Condition (C1) holds with $\gamma>1/a$. 
		 Under the null hypothesis, we have
		\begin{align*}
			V_n=:n\hat\sigma^{-2}_{Y|Z} T_n \overset{d}{\rightarrow} \chi^2(1).
		\end{align*}
  {Under local alternative hypotheses $H_{1n}: g(X)=\delta(X)/\sqrt n$ with $0<c<{\rm E}[\delta^2(X)]<C<\infty$, 
  \begin{align*}	V_n=n\hat\sigma^{-2}_{Y|Z} T_n \overset{d}{\rightarrow} \chi^2(1)+\sigma^{-2}_{Y|Z}{\rm E}[\delta^2(X)].		\end{align*}}
 Further assume ${\rm E}\left[\left\{m(X)-h(Z)\right\}^4\right]<\infty$.	{Under the fixed alternative hypothesis},
		\begin{align*}
			\sqrt{n}\left(T_n-{\rm E}[\left\{m(X)-h(Z)\right\}^2]\right) \overset{d}{\rightarrow} N\left(0,{\rm Var}\left[\{m(X)-h(Z)\}^2\right]\right).
		\end{align*}
\end{theorem}

Theorem \ref{theorem1} demonstrates that $T_n$ is $n$-consistent under the null hypothesis and the asymptotic null distribution is a standard chi-squared distribution, which provides a computationally efficient way to compute the p-values without any resampling procedures.  {Under the fixed alternative hypothesis, $T_n$ is root-$n$-consistent and  asymptotically normal.} {It also reveals that the proposed test has non-trivial power against the local alternative hypothesis which converges to the null at the root-$n$ rate. Although the root-$n$ rate is generally slower than the root-$N$ rate, it is very close to the root-$N$ rate under parametric regression models where we can take $a=1$ and $\gamma$ can be very close to 1.}
The condition $\gamma>1/a$ reveals the relative roles of the sample size $n_1$ used to estimate $g(X)$. If $g(X)$ can be estimated accurately, then $a$ could be close to 1 and the sample size $n_1$ can be reduced. {However, it is  challenging to determine $\gamma$ from a theoretical perspective, since $a$ is generally unknown, often related to the smoothness of the conditional mean function and the dimensionality. Practically,  we suggest achieving unbalanced sample splitting by tuning the ratio  of the sample size  of $\mathcal D_1$ to the total sample size, i.e., $\xi=n_1/N$. We  provide an adaptive procedure for selecting proper $\xi$ in   section \ref{sec2.2}, through which Type I error can be well controlled without much loss of power.}

{Now suppose that $\E[\{\hat g_{\D_1}(X)-\tilde g(X)\}^2]=O(n^{-a}_1)$. Here $\tilde g(X)$ may not be equal to the true function $g(X)$. As long as $\E[\tilde g^2(X)]>c_g$ for some positive constant $c_g$ under the alternative hypothesis, we have $V_n=nT_n/\hat\sigma_{Y|Z}^2=\sigma_{Y|Z}^{-2}
\sum_{i\in\D_2}\left[\tilde g(X_i)-n^{-1}\sum_{j\in \D_2}\left\{Y_j- h(Z_j)\right\}\right]^2+o_p(1)$ and thus $V_n\rightarrow \infty$. That is, we still can detect the alternative hypothesis even when $\hat g_{\D_1}(X)$ may not be a consistent estimator of $g(X)$.}

{We further note that {\color{blue}Theorem \ref{theorem1}} can be strengthened to uniform results. If we enhance the condition (C1) to be valid for all distributions $P$'s of $(X, Y)$ and {assume that $c_Y^\prime<\E|Y-h(Z)|^{2+\tau}<C_Y^\prime$ for all $P$'s with $\tau>0$ and $c_Y^\prime$, $C_Y^\prime$ being positive constants.} Then, from the Lemmas 18-20 in \cite{shah2020hardness} (see also 
Lemmas 16-17 in \cite{lundborg2022projected}), we can show that under the null hypothesis, $V_n$ converges to chi-squared distribution uniformly and the test can uniformly control the nominal level $\alpha$.

From Theorem \ref{theorem1}, in terms of ${\rm E}[g^2(X)]$, our procedure  can detect local alternative hypothesis which converges to the null at $n$ rate. With assumption that $1/2<a<1$ as in condition (C2), the convergence rate is faster than root-$N$ and slower than $N$ rate.  
While from \cite{lundborg2022projected}, the convergence rates of \cite{williamson2023general} and \cite{dai2022significance} are generally root-$N$. Although the procedure in \cite{lundborg2022projected} can achieve the fastest possible $N$ rate, the theorem is obtained with fixed dimension $W$ under linear regression models. Furthermore, compared with \cite{lundborg2022projected}, we establish the asymptotic distribution of the proposed test statistic under alternative hypothesis within the model-free framework. 

{In addition, we comment that a larger training sample could have a negative impact on the statistical power for our testing procedure. This is the price we have to pay for dealing with the estimation errors of machine learning methods and the critical degenerate limiting null distribution issue. Similarly, \cite{dai2022significance} also required larger training sample to make their bias-sd-ratio $T_2$ negligible. 
When the alternatives are believed to be complex, the use of machine learning methods including DNN can be helpful to estimate the conditional means accurately. If the relationship between the response and the covariates is believed to be parametric, parametric methods should be used and then it is not necessary to use the larger training sample. The loss of power associated with performing a test on a small testing sample could be offset by efficient estimation of regression functions. In the next session, we will also discuss some procedures for power enhancement. 


\subsection{Power enhancement and algorithm stability}\label{sec2.2}
We first present our pMIT procedure in Algorithm \ref{single1} based on single data splitting. 
{{\small \begin{algorithm}[htb]
		\caption{The pMIT procedure based single data splitting}
		\label{single1}
		\begin{algorithmic}
			\STATE \textbf{Step 1.}  For a given splitting ratio $\xi$, i.e. $\xi=n_1/N$, construct the pMIT test statistic $T_n$ in equation (\ref{pmittest})  and obtain $p$-value according to Theorem \ref{theorem1}, denoted as $p_0$.
			
			\STATE \textbf{Step 2.} Reject the null hypothesis if $p_0<\alpha$, where $\alpha$ is the pre-determined significance level.
			
		\end{algorithmic}
	
\end{algorithm}}

In the following, we make some improvements to enhance the empirical testing power.
Inspired by the seminal idea of \cite{fan2015power}, we first introduce a power-enhanced version of $V_n$, $V_n^*=V_n+\tau\sum_{i\in\D_2}\hat g^2_{\D_1}(X_i)$ with $\tau>0$. As discussed before under $H_0$, $\sum_{i\in\D_2}\hat g^2_{\D_1}(X_i)$ is asymptotically negligible and thus $V_n^*$ still converges to $\chi^2(1)$ under $H_0$, which implies that the size can be controlled asymptotically. While under the local alternative hypothesis $H_{1n}: g(X)=\delta(X)/\sqrt n$,  $\sum_{i\in\D_2}\hat g^2_{\D_1}(X_i)\rightarrow \E[\delta^2(X)]$ and thus $V_n^*\overset{d}{\rightarrow} \chi^2(1)+\sigma^{-2}_{Y|Z}{\rm E}[\delta^2(X)]+\tau {\rm E}[\delta^2(X)].$ Since $V_n^*\geq V_n$, the power is always enhanced. It reveals that the degeneracy issue can be a blessing in terms of power. Also the same idea can be applied to other recently developed test statistics.  How to adaptively choose $\tau$ warrants further investigation. Selecting an appropriate value of $\tau$ involves a trade-off between size control and power enhancement. With larger $\tau$, the power is larger but the empirical size faces the risk of being distorted. In practice, we simply take $\tau=1$.

			
			
			
				
 
Next, we discuss how to select the sample splitting ratio $\xi$ to balance the estimation bias and the test efficiency in Algorithm \ref{single1}, which can be viewed as the trade-off between type I and type II errors. {For the selection of $\xi$, \cite{dai2022significance} provided an easy-to-use formula and also a novel data-adaptive procedure. } 
Inspired by \cite{dai2022significance}, we also consider a data-adaptive splitting procedure and select optimal $\xi$ by controlling the estimated Type I error on permutation datasets. To be specific, we first randomize the covariates $W$ and obtain the corresponding $p$-values. After that, we estimate the Type I error by using the proportions of rejections over $M$ permutations, and select the value of $\xi$ that can control the estimated Type I error under some predetermined significance level. This procedure is presented in Algorithm \ref{Adaptive}. In the process of searching optimal $\xi$ in the candidate set $\Xi$, it stops once the criterion $\widehat{{\rm Err}}(\xi)\leq \alpha$ is met in practice to enjoy the computational efficiency.  
{\small \begin{algorithm}[h!]
		\caption{A data-adaptive procedure to select the sample splitting ratio $\xi$}
		\label{Adaptive}
		\begin{algorithmic}
			\STATE \textbf{Step 1.}  For a given splitting ratio $\xi$ in the candidate sets $\Xi$ (e.g. $\Xi=\{(k-1)/k: 2\leq k\leq \kappa\}$, $\kappa=10$), shuffle the covariates   $W$, and then apply Algorithm \ref{single1} with the splitting ratio $\xi$ to the permutation data and obtain the $p$-value, denoted as $p_1$.
			
			\STATE \textbf{Step 2.} Repeat Step 1 $M$ times (e.g., $M=200$). Obtain the set of $p$-values $\{p_1,p_2,\cdots,p_M\}$ corresponding to the ratio $\xi$. The estimated Type I error is given by
			\begin{align*}
				\widehat{{\rm Err}}(\xi) = M^{-1}\sum_{m=1}^{M}\mathbf{1}\left(p_m\leq \alpha\right).
			\end{align*}
			
			\STATE \textbf{Step 3.} Set $\xi^{\star}$ as the smallest value that
			controls the estimated Type I error, 
			\begin{equation*}
				\xi^{\star}=\min_{\xi\in\Xi}\left\{\widehat{{\rm Err}}(\xi)\leq \alpha\right\}.
			\end{equation*}
			
		\end{algorithmic}
\end{algorithm}}
\par Finally, the testing performance of the pMIT procedure outlined in Algorithm \ref{single1} may be affected by the additional randomness of the data splitting. To improve the algorithm stability, we introduce an ensemble testing procedure $\rm pMIT_M$ based on multiple data splittings in Algorithm \ref{multiple}. Actually we aggregate $p$-values from different sample splits to obtain 
an adjusted $p$-value $Q^*$. The $p$-value aggregation is an important problem in statistics. For recent theoretical investigations, see for instance \cite{diciccio2020exact}, \cite{vovk2020combining}, and \cite{choi2023averaging}. Instead of combining p-values, \cite{2023rank} studied the combination of test statistics from data splitting and made a deep theoretical investigation.

{\small \begin{algorithm}[htb]
	\caption{The pMIT procedure based multiple data splitting}
	\label{multiple}
	\begin{algorithmic}
		\STATE \textbf{Step 1.} Apply Algorithm \ref{single1} via a single random data splitting to the original data and obtain the $p$-value, denoted as $p_1$.
		
		\STATE \textbf{Step 2.} Repeat Step 1 $B$ times. Obtain the set of $p$-values $\{p_1,p_2,\cdots,p_B\}$.
		
		\STATE \textbf{Step 3.} Calculate the adjusted $p$-value $Q^*$ by aggregating the above $p$-values $\{p_1,p_2,\cdots,p_B\}$ in Step 2. (Note: different p-value aggregation methods are specified in the simulation section.)

		\STATE \textbf{Step 4.} Reject $H_0$ at a significance level $\alpha$ if $Q^\ast<\alpha$.		
	\end{algorithmic}
\end{algorithm}}

}

\section{Partial Generalized Measure of Correlation (pGMC)}
\subsection{Definition and properties}
The rejection of $H_0$ implies that the sub-vector of covariates $W$ is still able to add some contributions to the conditional mean of $Y$ after controlling for the nonlinear effect of $Z$. Then, it is of great interest to quantify the partial dependence between $W$ and $Y$ given $Z$. In this section, we propose a new partial dependence measure, called partial Generalized Measure of Correlation (pGMC), and derive its theoretical properties and confidence interval.

To quantify the partial dependence between $W$ and $Y$ given $Z$, a commonly used method is partial correlation coefficient, defined as
$$\rho(Y,W|Z)=\frac{\E[\{Y-\E(Y|Z)\}\{W-\E(W|Z)\}]}{\sqrt{\E[\{Y-\E(Y|Z)\}^2]}\sqrt{\E[\{W-\E(W|Z)\}^2]}},$$
which equals the Pearson correlation coefficient between
the errors $Y-\E(Y|Z)$ and $W-\E(W|Z)$. It only measures the linear correlation between $Y-\E(Y|Z)$ and $W-\E(W|Z)$.
It can be zero even there exists explicit partial dependence between $Y$ and $W$ given $Z$. For example, let $W$ and
$Z$ be two independent  standard normal random variables and set $Y=Z+W^2$. In this case, $Y$ and $W$ are perfectly dependent given $Z$ but $\rho(Y,W|Z)=0$. To measure the nonlinear dependence for the conditional mean of $Y$ given $W$ adjusting for $Z$,
we consider the following conditional variance decomposition
\begin{align*}
	\mbox{Var}(Y|Z)=\E\left\{\mbox{Var}(Y|X)|Z\right\}+\mbox{Var}\left\{\E(Y|X)|Z\right\}.
\end{align*}
It motivates us to consider the following partial measure
\begin{equation}\label{pGMC}
	 r^2(Y,W|Z)=\frac{\E[\mbox{Var}\{\E(Y|X)|Z\}]}{\E[\mbox{Var}(Y|Z)]}=\frac{\E[\{\E(Y|X)-\E(Y|Z)\}^2]}{\E[\{Y-\E(Y|Z)\}^2]}=\frac{\E[\{m(X)-h(Z)\}^2]}{\E[\{Y-h(Z)\}^2]},
\end{equation}
which can be interpreted as the explained variance of $Y$ by $W$ given $Z$ in a model-free framework. To make $r^2(Y, W|Z)$ well defined, it is necessary to assume that the denominator of (\ref{pGMC}) is nonzero. That is, $Y$ is not almost surely equal to a measurable function of $Z$. Otherwise, given $Z$, $Y$ is a constant almost surely and it makes no sense to discuss the contribution of $W$ to the conditional mean of $Y$.
When there is no conditional covariates set $Z$, $r^2(Y,W|Z)$ reduces to  $\mbox{GMC}(Y|X)=\mbox{Var}\left\{\E(Y|X)\right\}/\mbox{Var}(Y)$, the Generalized Measure of Correlation (GMC) proposed by \cite{zheng2012generalized}. 
Thus, the pGMC can be considered as an extension of the GMC to measure the partial dependence between $W$ and $Y$ given $Z$.

Next, we discuss more properties of $r^2(Y,Z|X)$ in the following proposition.

\begin{proposition}\label{prop1} The partial Generalized Measure of Correlation (pGMC) $r^2(Y,W|Z)$ defined in (\ref{pGMC}) satisfies that
	\begin{itemize}
		\item[1.] $r^2(Y,W|Z)$ is an indicator belonging to $[0,1]$. That is, $0\leq r^2(Y,W|Z)\leq 1.$
		\item[2.] $r^2(Y,W|Z)=0$ if and only if ${\rm E}(Y|X)={\rm E}(Y|Z),$ almost surely. That
		is, given $Z$, $W$ does not add further information about the
		conditional mean of $Y$.
		\item[3.] $r^2(Y,W|Z)=1$ if and only if $Y={\rm E}(Y|X),$ almost surely. That is, $Y$ is almost surely equal to a measurable function of $W$ given $Z$.
		\item[4.] The relation of $r^2(Y,W|Z)$ and the partial correlation coefficient $\rho(Y,W|Z)$ satisfies that  $r^2(Y,W|Z)\geq \rho^2(Y,W|Z)$ and  $\rho(Y,W|Z)\neq 0$ implies that $r^2(Y,W|Z)\neq 0$. If $r^2(Y,W|Z)=0$, then $\rho^2(Y,W|Z)=0$; If $\rho(Y,W|Z)=\pm 1$, then $r^2(Y,W|Z)=1$.
		\item[5.] The relation of $r^2(Y,W|Z)$ and ${\rm GMC}(Y|W)$ satisfies that
		$$r^2(Y,W|Z)=\frac{{\rm Var}\left\{{\rm E}(Y|X)\right\}-{\rm Var}\left\{{\rm E}(Y|Z)\right\}}{{\rm Var}(Y)-{\rm Var}\left\{{\rm E}(Y|Z)\right\}}=\frac{{\rm GMC}(Y|X)-{\rm GMC}(Y|Z)}{1-{\rm GMC}(Y|Z)}.$$
		If ${\rm E}(Y|Z)={\rm E}(Y)$, then $r^2(Y,W|Z)={\rm GMC}(Y|X)={\rm GMC}(Y|W).$
		\item[6.] $r^2(Y,W|Z)$ can also be rewritten as
		$$r^2(Y,W|Z)=\frac{{\rm E}\left[\{Y-h(Z)\}^2\right]-{\rm E}\left[\{Y-m(X)\}^2\right]}{{\rm E}\left[\{Y-h(Z)\}^2\right]}=1-\frac{{\rm E}\left[\{Y-m(X)\}^2\right]}{{\rm E}\left[\{Y-h(Z)\}^2\right]}.$$
		\item[7.] Suppose $Z_s$ is a subset of $Z$ and $X=(Z_s^\top, W_s^\top)^\top$, then  
		${\rm GMC}(Y|X)\geq r^2(Y,W_s|Z_s)\geq r^2(Y,W|Z).$
	\end{itemize}
\end{proposition}

In Proposition \ref{prop1}, properties (1)-(3) demonstrate that the new partial dependence $r^2(Y,W|Z)$ characterizes nonlinearly partial dependence. $r^2(Y,W|Z)$  is 0 if and only if $Y$ and $W$ are conditionally mean independent given $Z$, and is 1 if and only if $Y$ is equal to a measurable function of $W$ given $Z$. For example, for $Y=Z+W^2$ where $W$ and $Z$ be two independent standard normal random variables, $r^2(Y,W|Z)=1$.
In property (4), $r^2(Y,W|Z)\geq \rho^2(Y,W|Z)$ further indicates that $r^2(Y,W|Z)$ is capable of revealing more conditional association information between $Y$ and $W$ beyond the linear relation.  The relationship between the new partial dependence and the GMC in \cite{zheng2012generalized} is established in property (5). As the GMC is viewed as a generalization of the classical $R^2$ measurement to a nonparametric model, the new partial dependence can then be viewed as the relative difference in the population $R^2$ obtained using the full set of covariates $X$ as compared to the subset of covariates $Z$ only. Moreover, when the conditional mean of $Y$ does not depend on $Z$, the new partial dependence $r^2(Y,W|Z)$ then reduces to $\mbox{GMC}(Y|X)$. Thus $r^2(Y,W|Z)$ is a natural extension of $\mbox{GMC}(Y|X)$ for partial mean dependence. From property (6), $r^2(Y,W|Z)$ measures the relative reduction of population residual sum of squares with additional $W$ in the model. The last property (7) means that when the conditional information is more, the remaining variables provide less additional information about the conditional mean of the response. On the other hand, the partial dependence is always bounded by the $\mbox{GMC}(Y|X)$.

\begin{remark} \label{remarkpgmc}
	The pGMC $r^2(Y,W|Z)$ is related with the novel conditional dependence measure proposed by \cite{azadkia2021simple}. Their measure is defined as
	$$\xi(Y,W|Z)=\frac{\int {\rm E}[{\rm Var}\{{\rm E}(I(Y\geq y)|X)|Z\}]dF_Y(y)}{\int {\rm E}[{\rm {\rm Var}}\{I(Y\geq y)|Z\}]dF_Y(y)},$$
	where $F_Y(y)$ is the distribution function of $Y$. By comparing the formulas for $r^2(Y,W|Z)$ and $\xi(Y,W|Z)$, $r^2(Y,W|Z)$ can be considered as  the conditional mean version of $\xi(Y,W|Z)$. As \cite{cook2002dimension} and \cite{shao2014martingale}, conditional mean is of primary interest in many applications such as regression problems. The pGMC provides a nonparametric way to measure the additional contribution of $W$ for the conditional mean of $Y$ given $Z$.

	We remark that \cite{williamson2021nonparametric} introduced
	an appealing variable importance measure
	$$\phi(Y,W|Z)=\frac{{\rm E}[\{m(X)-h(Z)\}^2]}{{\rm E}[\{Y-{\rm E}(Y)\}^2]},$$
	which is indeed the difference between ${\rm GMC}(Y|X)$ and ${\rm GMC}(Y|Z)$. It can measure the importance of $W$ for $Y$. {However, it cannot be a correlation measure for the partial mean dependence of $Y$ given $W$ conditional on $Z$.} Because $\phi(Y,W|Z)=1$  if and only if ${\rm GMC}(Y|X)=1$, i.e. $Y={\rm E}(Y|X)$, and
	${\rm GMC}(Y|Z)=0$, i.e. ${\rm E}(Y|Z)={\rm E}(Y)$, which requires that the mean of $Y$ doesn't depend on $Z$. Thus, its interpretation of $\phi(Y,W|Z)=1$ is unclear. 
{The pGMC $r^2(Y,W|Z)$ shares the same numerator as $\phi(Y,W|Z)$, which measures the  difference between two conditional means $m(X)$ and $h(Z)$.
 The difference between $\phi(Y,W|Z)$ and the pGMC is the normalization term in the denominator. The pGMC adopts  ${\rm E}[\{Y-h(Z)\}^2]$ as the normalization term, which  makes the pGMC be a suitable partial mean dependence measurement with clear interpretation. The pGMC $r^2(Y,W|Z)=1$ if and only if $Y$ is equal to a measurable function of $W$ given $Z$. More interesting properties of the pGMC as a partial mean dependence measure are also established in Proposition \ref{prop1}.  
  These distinct attributes position our pGMC as a valuable augmentation to the current body of literature on partial mean dependence measures.}
\end{remark}

\subsection{Estimation and confidence interval}
Next, we estimate the pGMC $r^2(Y,W|Z)$ via data splitting. We split the data randomly into two independent parts with equal sample size, denoted by $\D_1$ and $\D_2$, and estimate the conditional means $h(Z)$ and $m(X)$ using $\D_1$. According to the representation of $r^2(Y,W|Z)$ in property (6), we estimate its numerator
$\E\left[\{Y-h(Z)\}^2\right]-\E\left[\{Y-m(X)\}^2\right]$ by the following statistic using $\D_2$,
\begin{align*}
	R_n=\frac{1}{n}\sum_{i\in \D_2}\left\{Y_i-\hat{h}_{\D_1}(Z_i)\right\}^2-\frac{1}{n}\sum_{i\in \D_2}\left\{Y_i-\hat{m}_{\D_1}(X_i)\right\}^2.
\end{align*}
To improve the estimation efficiency, we use the cross-fitting method to obtain another estimator
$R_n^\prime$ similarly by swapping the roles of $\D_1$ and $\D_2$.
Then, we define $R_n^\ast=(R_n+R_n^\prime)/2$ to estimate the numerator of $r^2(Y,W|Z)$.
We remark that the cross-fitting method  has been shown to improve the estimation efficiency asymptotically in \cite{fan2012variance}, \cite{chernozhukov2018double} and \cite{vansteelandt2022assumption}. Similarly, we can estimate the denominator $\E\left[\left\{Y-h(Z)\right\}^2\right]$ of $r^2(Y,W|Z)$ by $\widehat{\sigma}^{2*}_{Y|Z}=\left(\widehat{\sigma}^2_{Y|Z}+\widetilde{\sigma}^{2}_{Y|Z}\right)/2$, where $\widehat{\sigma}^2_{Y|Z}={n}^{-1}\sum_{i\in \D_2}\left\{Y_i-\hat h_{\D_1}(X_i)\right\}^2,
$ and $\widetilde{\sigma}^2_{Y|Z}$ is similarly defined by swapping the role of $\D_1$ and $\D_2$.
Thus, the pGMC $r^2(Y,W|Z)$ can be estimated by
\begin{align*}
	\hat r^2(Y,W|Z)=\frac{R_n^\ast}{\widehat{\sigma}^{2*}_{Y|Z}}.
\end{align*}

It is worth noting that the balanced data splitting and the cross-fitting method is used to estimate the pGMC, which is different from the unbalanced data splitting for the significance testing procedure in the previous section. This is because, when the partial dependence between $W$ and $Y$ given $Z$ truly exists, i.e. $r^2(Y,W|Z)\neq 0$, we can use the balanced data splitting to improve the estimation efficiency under the following condition (C2).
\begin{itemize}
	\item[(C2)]
	$\E\left[\left\{\hat m_{\D_1}(X)-m(X)\right\}^2\right]=o(n^{-1/2}_1)$ and $\E\left[\sigma^2(X)\left\{\hat m_{\D_1}(X)-m(X)\right\}^2\right]=o(1)$, where
	$\epsilon=Y-m(X)$ and $\sigma^2(X)=\E(\epsilon^2|X)$;
	$\E\left\{\hat h_{\D_1}(Z)-h(Z)\right\}^2=o(n^{-1/2}_1)$ and\\ $\E\left[\delta^2(X)\left\{\hat h_{\D_1}(Z)-h(Z)\right\}^2\right]=o(1)$, where
	$\eta=Y-h(Z)$ and $\delta^2(X)=\E\left(\eta^2|X\right)$.
\end{itemize}
Here $n_1=n_2=n$. 
 {The assumptions $\E\left[\left\{\hat m_{\D_1}(X)-m(X)\right\}^2\right]=o(n^{-1/2})$ and $\E\left[\left\{\hat h_{\D_1}(Z)-h(Z)\right\}^2\right]=o(n^{-1/2})$ in (C2) are also adopted in \cite{chernozhukov2018double} (equation 3.8), \cite{williamson2021nonparametric} (assumption A1) and in \cite{vansteelandt2022assumption} (assumptions in Theorem 2). For a full discussion about the estimator of nuisance functions, see \cite{chernozhukov2018double}. \cite{chernozhukov2018double} showed that 
the $l_1$-penalized and related methods in various sparse models can satisfy the above condition.  For the DNN, \cite{bauer2019deep} showed that ${\rm E}\left[\left\{\hat m_{\D_1}(X)-m(X)\right\}^2\right]=O((\log n)^3n^{-2q/(2q+d^*)})$, where $q$ is the smoothness parameter and $d^*$ is the intrinsic dimensionality. For other recent developments, see also  \cite{schmidt2020nonparametric}  and \cite{kohler2021rate}. While the other assumptions $\E\left[\sigma^2(X)\left\{\hat m_{\D_1}(X)-m(X)\right\}^2\right]=o(1)$ and $\E\left[\delta^2(X)\left\{\hat h_{\D_1}(Z)-h(Z)\right\}^2\right]=o(1)$ follow if we assume that the conditional variance functions $\sigma^2(X)$ and $\delta^2(X)$ are bounded above.} 

Next, we establish the asymptotic normality for the $\hat r^2(Y,W|Z)$ in Theorem \ref{asymnorm}.
\begin{theorem} \label{asymnorm}
	Suppose Conditions (C2) and ${\rm E}\left[\left\{Y-h(Z)\right\}^4\right]<\infty$ are satisfied. When $r^2(Y,W|Z)\neq 0$, we have
	\begin{equation} \label{asymnormeqn}
		\sqrt{N}\left\{\hat r^2(Y,W|Z)-r^2(Y,W|Z)\right\} \overset{d}{\rightarrow} N\left(0,{\rm Var}(\Phi)\right),
	\end{equation}
	where $$\Phi = \frac{\left\{m(X)-h(Z)\right\}^2+2\epsilon\left\{m(X)-h(Z)\right\}}{\sigma_{Y|Z}^2}
	- \frac{{\rm E}\left[\left\{m(X)-h(Z)\right\}^2\right]\eta^2}{\sigma_{Y|Z}^4}.$$
\end{theorem}

As discussed in Remark \ref{remarkpgmc}, $r^2(Y,W|Z)$ is the conditional mean version of $\xi(Y,W|Z)$ introduced in \cite{azadkia2021simple}. However, from the above theorem, one can see that the asymptotic properties of two estimators are different.  \cite{azadkia2021simple} only derived the convergence rate
of the estimator for $\xi(Y,W|Z)$, which is slightly larger than $N^{-1/p}$ and depends strongly on the dimension of both $Z$ and $W$. Our convergence rate $N^{-1/2}$ is significantly faster than it.  Besides, we also establish the asymptotic normality which paves a way to derive the confidence interval for $r^2(Y,W|Z)$.

Next, we estimate the asymptotic variance in (\ref{asymnormeqn}) and  derive the confidence interval for $r^2(Y,W|Z)$. Define
\begin{equation}\label{mhhat}
	\hat m(X_i)=\left\{
	\begin{array}{rcl}
		\hat m_{\D_1}(X_i), & & {i \in \D_2};\\
		\hat m_{\D_2}(X_i), & & {i \in \D_1},\\
	\end{array} \right.
	~~~
	\hat h(Z_i)=\left\{
	\begin{array}{rcl}
		\hat h_{\D_1}(Z_i), & & {i \in \D_2};\\
		\hat h_{\D_2}(Z_i), & & {i \in \D_1},\\
	\end{array} \right.
\end{equation}
$\hat\epsilon_i=Y_i-\hat m(X_i)$ and $\hat\eta_i=Y_i-\hat h(Z_i)$.
A natural plug-in estimator of $\mbox{Var}(\Phi)$ is given by
\begin{equation}\label{phihat}
	\widehat{\mbox{Var}}(\Phi)=\frac{1}{N}\sum_{i=1}^{N}\hat\Phi_i^2, \end{equation}
where $\hat\Phi_i$ is any consistent estimator of $\Phi_i$. For example, $\hat\Phi_i$ might be taken as $\Phi_i$ with $\sigma_{Y|Z}^2$, $h(Z_i)$, $m(X_i)$, $\epsilon_i$, $\eta_i$, and ${\rm E}\{m(X)-h(Z)\}^2$ replaced by $\hat\sigma_{Y|Z}^{2*}$, $\hat h(Z_i)$, $\hat m(X_i)$, $\hat{\epsilon}_i$, $\hat \eta_i$, and $R_n^\ast$, respectively.
Hence, a confidence interval at the confidence level of $1-\alpha$ for $r^2(Y,W|Z)$ is
\begin{align*} 
	\left( \frac{R_n^\ast}{\hat{\sigma}_{Y|Z}^{2*}}-z_{\alpha/2}\cdot\sqrt{ \frac{\widehat{\mbox{Var}}(\Phi)}{N}},\,\,
	\frac{R_n^\ast}{\hat{\sigma}_{Y|Z}^{2*}}+z_{\alpha/2}\cdot\sqrt{ \frac{\widehat{\mbox{Var}}(\Phi)}{N}}
	\right),
\end{align*}
where $z_{\alpha/2}$ is the $\alpha/2$ upper-tailed  critical value of the standard normal distribution.

\section{Special case: Conditional mean independence Test}
As an important special case when there is no conditioning random object $Z$, the original pMIT test  (\ref{partialmeantest}) becomes
\begin{equation*}	\widetilde{H}_0:\,\,\,\E(Y|X)=\E(Y), \mbox{~v.s.~~~} \widetilde{H}_1:\,\,\,\E(Y|X)\ne \E(Y).
\end{equation*}
$\widetilde{H}_0$  indicates that the conditional mean of $Y$ given $X$ does not depend on $X$. It is  the fundamental testing problem to check the overall significance of covariates $X$ for modeling the mean of the response $Y$. The test has to be conducted before any regression modeling.
Similar to the pMIT test statistic (\ref{pmittest}), we consider the test statistic via comparing the estimators of $\E(Y|X)$ and $\E(Y)$ based on data splitting and machine learning methods,
\begin{align*}
	T_{1n} = \frac{1}{n}\sum_{i\in \D_2}\left\{\hat m_{\D_1}(X_i)-\bar{Y}_{\D_2}\right\}^2,
\end{align*}
where $\bar{Y}_{\D_2}=n^{-1}\sum_{j\in \D_2}Y_j$ is the average of $Y$ over the second part of data. 
With unbalanced sample splitting strategy, $\sum_{i\in\D_2}\left\{\hat m_{\D_1}(X_i)-\E(Y)\right\}^2$ becomes negligible compared with $(\bar{Y}_{\D_2}-\E(Y))^2$. Denote $\sigma_Y^2:=\mbox{Var}(Y)$ and $\hat\sigma_Y^2=N^{-1}\sum_{i=1}^N\left(Y_i-N^{-1}\sum_{i=1}^N Y_i\right)^2$. We impose similar estimation error condition of the conditional mean of $Y$ given $X$ via machine learning methods.
\begin{itemize}
	\item[(C1')] $\E\left[\left\{\hat m_{\D_1}(X)-m(X)\right\}^2\right]\leq c n_1^{-a}=c n^{-\gamma a}$ for some $0<a<1$.
\end{itemize}
{The above condition (C1') is the first part of the condition (C1) with $m(X)$ and $\hat m_{\D_1}(X)$.} The asymptotic distributions of $T_{1n}$ are established in the following corollary.
\begin{corollary} \label{coro1}
	{Assume that ${\rm E}(Y^2)<\infty$ and Condition (C1') holds with $\gamma>1/a$.}
	When the null hypothesis holds, we have
	\begin{align*}
		n\hat\sigma_Y^{-2} T_{1n} \overset{d}{\rightarrow} \chi^2(1).
	\end{align*}
  {Under local alternative hypotheses $\widetilde{H}_{1n}: {\rm E}(Y|X)-{\rm E}(Y)=\Delta(X)/\sqrt n$ with $0<c<{\rm E}[\Delta^2(X)]<C<\infty$, 
  \begin{align*}	n\hat\sigma_Y^{-2} T_{1n}\overset{d}{\rightarrow} \chi^2(1)+\sigma^{-2}_{Y}{\rm E}[\Delta^2(X)].		\end{align*}}
	{Further, assume that ${\rm E}[m(X)^4]<\infty$.}
	When the fixed alternative hypothesis holds,
	\begin{align*}
		\sqrt{n}\left(T_{1n}-{\rm E}\left[\{m(X)-\mu\}^2\right]\right) \overset{d}{\rightarrow} N\left(0,{\rm Var}\left[\left\{m(X)-\mu\right\}^2\right]\right).
	\end{align*}
\end{corollary}

Corollary \ref{coro1} demonstrates that the asymptotic null distribution of the test statistic $T_{1n}$ after the standardization is the standard chi-squared distribution. Asymptotic distributions under local and fixed alternative hypotheses are also derived. 
We remark that the proposed significance test can be applied to high dimensional settings in a model-free framework. It is more appealing than the methods proposed by \cite{zhang2018conditional} and \cite{li2023testing}, which considered the sum-type test statistics over all marginal tests under a weaker but necessary null hypothesis (\ref{weaknull}). Because the summation of all marginal test statistics ignore the interactions among covariates and can fail for pure interaction models.
We will show this comparison by the simulations in the supplementary materials.
Moreover, we can apply the multiple data splitting to improve the algorithm stability. Since it is similar to Algorithm 3 as mentioned in Section 2, we omit the details of the algorithm here.

\section{Numerical Studies}\label{sec5}
In this section, we present some numerical results to assess the finite sample performance of our proposed model-free statistical inference methods. To conserve space, extensive additional simulation results are included in the supplementary materials.

{\subsection{Partial mean independence test}
We consider the following two examples including linear and nonlinear models.

\textbf{Example A1:}  We first consider a linear regression
\begin{equation}
	Y_i =  Z_{i1}+Z_{i2} +  \sum_{j=1}^{p_2}\theta_j W_{ij} + \epsilon_i \ \ \ i = 1,2,...,N,\nonumber
\end{equation}
where  $Z_i=\left(Z_{i1},\ldots,Z_{ip_1}\right)^\top$ and $W_i=\left(W_{i1},\ldots,W_{ip_2}\right)^\top$, and we set $p_1=p_2=25$. For each $i$, the covariate vector $X_i = \left(Z_i^\top, W_i^\top\right)^\top$ is generated from the multivariate normal distribution 
$N(0, \Sigma)$ with ${\Sigma}_{ij}=0.3^{|i-j|}$ for $i,j=1,2,\cdots,p$. 
The stochastic error term $\epsilon_i$ follows  the normal distribution $N(0,0.5^2)$. 
We consider three different settings of ${\theta}=(\theta_1,\ldots,\theta_{p_2})$: (1) the null hypothesis $H_0$, set $\theta_j=0$ for all $1\leq j\leq p_2$; (2) the sparse alternative $H_1$, the first two elements of ${\theta}$ are nonzeros with equal magnitude and  $\left \|{\theta}\right \|=1/\sqrt{2}$; (3) the dense alternative $H_1$,  each element of  ${\theta}$ is nonzero with equal magnitude  and  $\left \|{\theta}\right \|=1/\sqrt{2}$.

\textbf{Example A2:}  We consider a nonlinear regression
\begin{equation}
	Y_i = Z_{i1}+Z_{i2} +  \left(\sum_{j=1}^{p_2}\theta_j W_{ij}\right)^2 + \epsilon_i \ \ \ i = 1,2,...,N,\nonumber
\end{equation}
where  $X_i$ and $\epsilon_i$ are generated by the same setting as Example A1. 
The response $Y$ depends on $W$ via a quadratic term of a linear combination of $W$. (1) The null hypothesis $H_0$ holds when all $\theta_j=0$. 
We consider two slightly different settings of ${\theta}$ from Example A1: (2) the sparse  $H_1$, the first five elements of ${\theta}$ are nonzeros with equal magnitude and  $\left \|{\theta}\right \|=1/\sqrt{2}$; (3)
the dense  $H_1$,  the first half elements of  ${\theta}$ are nonzero with equal magnitude  and  $\left \|{\theta}\right \|=1/\sqrt{2}$.


In each simulation, we estimate the conditional mean functions using the XGBoost (eXtreme Gradient Boosting) procedure \citep{chen2016xgboost} based on the R package \textit{{xgboost}},  and deep neural network by employing the \textit{torch} module in Python.  
We set $\mathtt{eta}=0.1$ and $\mathtt{nround}=200$ as the default setting of the hyperparameters in XGBoost.
For the  neural network, we consider the fully connected neural network with depth $L=3$ and the width $\varpi=50$. We set the learning rate as $0.003$, the batchsize as $50$, and the maximum number of training epochs as $200$. The Adam optimizer is adopted for parameter optimizing.
In order to prevent under-fitting or over-fitting, it is vital to care about  the generalization ability of the model. Thus, we involve an early stopping strategy,  a useful practical technique to reduce the generalization error  efficiently and prevent over-fitting. It is a data-driven approach to determine the number of training epochs such that the model stops training when validation loss doesn't improve after a given patience (a few epochs).  In practice, we  use the testing data to validate the model.


Recall that we start with unbalanced single data splitting in Algorithm \ref{single1}. To evaluate the effect of the data splitting ratio, we consider different choices of $\xi=n_1/N=0.5,0.6,\ldots,0.9$ in numerical examples. We also implement the   easy-to-use formula provided in \cite{dai2022significance} with $N_0=0.1N$ defined therein, and the proposed adaptive data splitting procedure in Algorithm \ref{Adaptive}. From Table \ref{ratio-A1} in the Supplement, it is interesting to note that for balanced data splitting (the case of $\xi=0.5$), the empirical size of pMIT is hard to be controlled under $H_0$, which is consistent with our theoretical results in Theorem \ref{theorem1}. As $\xi$ becomes larger, the empirical sizes are getting closer to the significance level $0.05$, while the test loses powers due to the smaller sample size of the second part of the data. We find that both the easy-to-use formula and the data-adaptive splitting procedure work well, while the data-adaptive splitting procedure generally has higher powers than the easy-to-use formula.

Then, we compare different aggregation methods in the proposed ensemble partial mean independence test, including the multiple random splitting procedure in \cite{Meinshausen2009}, the Cauchy combination approach in \cite{liu2020cauchy}, and the Rank-transformed subsampling method in \cite{2023rank}. We summarize  these
combining methods in Table \ref{pvalue-combine} in the Supplement. In our simulation results, we find that the Rank-transformed subsampling method has the largest powers, while multiple random splitting procedure in \cite{Meinshausen2009} generally is conservative. However, the computation cost of the Rank-transformed subsampling method is generally heavy. Thus, we adopt the Cauchy combination approach for its computational efficiency and detection of the signal in the following simulations.

To proceed, we conduct simulation experiments to assess the finite sample performance of our proposed single data splitting procedure in Algorithm \ref{single1}, denoted as pMIT, the multiple data splitting ($B=10$) in Algorithm \ref{multiple}, denoted as $\rm pMIT_M$, the power enhanced procedure, denoted as $\rm pMITe$, and its multiple data splitting version, denoted as $\rm pMITe_{M}$. {The adaptive data splitting in Algorithm \ref{Adaptive} is adopted to determine the splitting ratio $\xi$.} For the purpose of comparison, we also implement the Algorithm 3 in \cite{williamson2023general}, which is implemented by the cv\_vim function from the R package \textit{vimp} {with  $K=2$, resulting in 4 folds}, denoted as vim;  the Projected Covariance Measure in Algorithm 1 of \cite{lundborg2022projected}  {with their practically recommended splitting ratio $\xi=1/2$}, denoted as pcm; the multiple splitting version in Algorithm 2 of \cite{lundborg2022projected} with $B = 6$ splits, denoted as $\rm pcm_M$; and the one-split test and the combined one-split test  in \cite{dai2022significance} {with their proposed adaptive splitting scheme}, denoted as DSP and $\rm DSP_M$, respectively.
For each experiment, we run 500 replications to compute the empirical sizes and powers.
The simulation results are presented in Tables \ref{ExampleA1}-\ref{ExampleA2}. Additional simulations are referred to the Supplement.

\begin{table}[htbp]
  \centering
  \caption{{\small Comparison of empirical sizes and powers of partial mean independence test using XGBoost (left) and DNN(right) in Example A1 at the significance of $\alpha=0.05$.}}
  \resizebox{\textwidth}{!}{
    \begin{tabular}{ccccccccccccccc}
    \toprule
          & N     & \multicolumn{2}{c}{pMIT} & \multicolumn{2}{c}{$\rm pMIT_M$} & \multicolumn{2}{c}{pMITe} & \multicolumn{2}{c}{$\rm pMITe_M$} & pcm   & $\rm pcm_M$ & vim   & DSP   & $\rm DSP_M$ \\
          &       & XGB & DNN   & XGB & DNN   & XGB & DNN   & XGB & DNN   &       &       &       &       &  \\
          \midrule
    \multirow{4}[1]{*}{$\rm H_0$} & 100   & 0.032  & 0.060  & 0.032  & 0.076  & 0.050  & 0.064  & 0.060  & 0.076  & 0.058  & 0.000  & 0.048  & 0.000  & 0.000  \\
          & 200   & 0.062  & 0.030  & 0.052  & 0.060  & 0.062  & 0.054  & 0.056  & 0.062  & 0.030  & 0.000  & 0.036  & 0.000  & 0.000  \\
          & 300   & 0.038  & 0.046  & 0.050  & 0.050  & 0.060  & 0.048  & 0.054  & 0.072  & 0.026  & 0.000  & 0.032  & 0.000  & 0.000  \\
          & 400   & 0.050  & 0.044  & 0.054  & 0.050  & 0.056  & 0.050  & 0.066  & 0.070  & 0.016  & 0.000  & 0.030  & 0.000  & 0.000  \\
    \midrule
    \multirow{4}[2]{*}{$\rm H_1$ sparse} & 100   & 0.110  & 0.136  & 0.146  & 0.212  & 0.284  & 0.186  & 0.538  & 0.250  & 0.138  & 0.008  & 0.308  & 0.060  & 0.090  \\
          & 200   & 0.442  & 0.492  & 0.926  & 0.886  & 0.740  & 0.522  & 0.994  & 0.908  & 0.684  & 0.794  & 0.880  & 0.416  & 0.690  \\
          & 300   & 0.822  & 0.738  & 1.000  & 1.000  & 0.984  & 0.764  & 1.000  & 1.000  & 0.966  & 1.000  & 0.962  & 0.852  & 1.000  \\
          & 400   & 0.984  & 0.818  & 1.000  & 1.000  & 1.000  & 0.828  & 1.000  & 1.000  & 0.998  & 1.000  & 1.000  & 0.994  & 1.000  \\
    \midrule
    \multirow{4}[2]{*}{$\rm H_1$ dense} & 100   & 0.076  & 0.228  & 0.068  & 0.410  & 0.246  & 0.316  & 0.470  & 0.496  & 0.126  & 0.000  & 0.244  & 0.264  & 0.384  \\
          & 200   & 0.210  & 0.674  & 0.272  & 0.982  & 0.554  & 0.710  & 0.930  & 0.996  & 0.430  & 0.280  & 0.662  & 0.808  & 1.000  \\
          & 300   & 0.492  & 0.896  & 0.732  & 1.000  & 0.866  & 0.908  & 0.998  & 1.000  & 0.836  & 0.956  & 0.908  & 0.994  & 1.000  \\
          & 400   & 0.804  & 0.938  & 0.974  & 1.000  & 0.986  & 0.952  & 1.000  & 1.000  & 0.976  & 1.000  & 0.974  & 1.000  & 1.000  \\
    \bottomrule
    \end{tabular}%
    }
  \label{ExampleA1}%
\end{table}%

\begin{table}[htbp]
  \centering
    \caption{{\small Comparison of empirical sizes and powers of partial mean independence test using XGBoost (left) and DNN (right) in Example A2 at the significance of $\alpha=0.05$. }}
    \resizebox{\textwidth}{!}{
    \begin{tabular}{ccccccccccccccc}
    \toprule
          & N     & \multicolumn{2}{c}{pMIT} & \multicolumn{2}{c}{$\rm pMIT_M$} & \multicolumn{2}{c}{pMITe} & \multicolumn{2}{c}{$\rm pMITe_M$} & pcm   & $\rm pcm_M$ & vim   & DSP   & $\rm DSP_M$ \\
          &       & XGB & DNN   & XGB & DNN   & XGB & DNN   & XGB & DNN   &       &       &       &       &  \\
    \midrule
    \multirow{4}[2]{*}{$\rm H_0$} & 100   & 0.034  & 0.044  & 0.032  & 0.036  & 0.042  & 0.046  & 0.054  & 0.050  & 0.058  & 0.000  & 0.048  & 0.000  & 0.000  \\
          & 200   & 0.054  & 0.046  & 0.036  & 0.054  & 0.060  & 0.060  & 0.048  & 0.076  & 0.024  & 0.000  & 0.036  & 0.000  & 0.000  \\
          & 300   & 0.048  & 0.060  & 0.054  & 0.036  & 0.052  & 0.060  & 0.056  & 0.074  & 0.018  & 0.000  & 0.058  & 0.000  & 0.000  \\
          & 400   & 0.048  & 0.038  & 0.060  & 0.056  & 0.056  & 0.052  & 0.062  & 0.076  & 0.028  & 0.000  & 0.050  & 0.000  & 0.000  \\
    \midrule
    \multirow{4}[2]{*}{$\rm H_1$ sparse} & 100   & 0.126  & 0.104  & 0.270  & 0.172  & 0.330  & 0.198  & 0.650  & 0.446  & 0.074  & 0.000  & 0.054  & 0.020  & 0.044  \\
          & 200   & 0.164  & 0.220  & 0.398  & 0.382  & 0.466  & 0.296  & 0.896  & 0.642  & 0.088  & 0.008  & 0.224  & 0.078  & 0.128  \\
          & 300   & 0.262  & 0.284  & 0.538  & 0.640  & 0.698  & 0.326  & 0.964  & 0.808  & 0.234  & 0.052  & 0.446  & 0.336  & 0.652  \\
          & 400   & 0.450  & 0.356  & 0.840  & 0.826  & 0.914  & 0.410  & 1.000  & 0.932  & 0.442  & 0.446  & 0.662  & 0.742  & 0.946  \\
    \midrule
    \multirow{4}[2]{*}{$\rm H_1$ dense} & 100   & 0.126  & 0.114  & 0.288  & 0.170  & 0.368  & 0.208  & 0.714  & 0.464  & 0.062  & 0.000  & 0.050  & 0.044  & 0.024  \\
          & 200   & 0.152  & 0.222  & 0.344  & 0.438  & 0.428  & 0.388  & 0.872  & 0.730  & 0.062  & 0.000  & 0.100  & 0.242  & 0.388  \\
          & 300   & 0.160  & 0.376  & 0.386  & 0.800  & 0.518  & 0.440  & 0.924  & 0.876  & 0.088  & 0.000  & 0.198  & 0.610  & 0.852  \\
          & 400   & 0.214  & 0.496  & 0.484  & 0.934  & 0.652  & 0.552  & 0.986  & 0.956  & 0.140  & 0.012  & 0.294  & 0.880  & 1.000  \\
    \bottomrule
    \end{tabular}%
    }
  \label{ExampleA2}%
\end{table}%

According to the simulation results, we have the following findings. Firstly, the empirical sizes of our proposed tests  and also the vim procedure are closely around the significance level $0.05$, while the empirical sizes of $\rm pcm_M$, DSP and $\rm DSP_M$ are very small. 
Secondly, power enhancement strategy and multiple data splitting are necessarily effective to improve empirical powers of the proposed pMIT. Thus, we recommend the multiple data splitting version of the power enhanced test statistic, $\rm pMITe_{M}$. Thirdly, the proposed methods are strong alternatives to existing tests in most cases. The empirical powers of our procedure with power enhancement strategy and multiple data splitting are similar or even higher than the other procedures. For instance, in Example A2, the pcm and vim procedures do not have high powers for dense alternative hypotheses. 

In addition, we also conduct numerical simulations to assess the impact of  different network structure (the depth $L$ and the width $\varpi$).  The corresponding results are provided in Table \ref{DNN_different}, indicating that  the performance of the proposed methods is not sensitive to the network structure. Furthermore, we also check the robustness of the proposed methods under the case of heterogeneous error via simulations. Please see Section A.2 in the Supplement. }

\subsection{Confidence interval for pGMC}
In this subsection, we conduct numerical simulations  to evaluate the empirical performance of the  proposed confidence interval for the pGMC. We consider two models in the following.

\textbf{Example B1:}
\begin{equation}
	Y_i =  Z_i^\top\beta + W_i^\top\theta + \epsilon_i \ \ \ i = 1,2,...,N, \nonumber
\end{equation}
where $Z_i$ is a $[p/2]$-dimensional vector. The covariates $Z_i$ and $W_i$ are independent, and  generated from  
$Z_i,W_i\sim N(0,\Sigma)$ with $\Sigma_{ij}=\rho^{|i-j|}$ for $i,j=1,2,\cdots,p$ with $\rho=0.5$, respectively.
Besides, only the first three elements of $\beta$ and $\theta$ are nonzeros with equal magnitude such that $\left \|\beta  \right \|= \left \|\theta  \right \|=1$. The error term follows the standard normal distribution.

\textbf{Example B2:}
\begin{equation}
	Y_i = Z_{i1} + 2\sin\left(\frac{W_{i1}}{2}\right) + \epsilon_i \ \ \ i = 1,2,...,N.\nonumber
\end{equation}
where $Z_i$ is a $[p/2]$-dimensional vector. $Z_{i1}$ and $W_{i1}$ denote the first element of the covariates $Z_i$ and $W_i$, respectively.  The other settings are the same as Example B1.

To make inference for the pGMC of the simulated data, we build a fully connected neural network with the default settings and involve distance correlation based feature screening techniques \citep{li2012feature} before training. To be more specific, we first apply the distance correlation based feature screening approach to the training data in order to reduce dimension of the predictors. Then we use the reduced predictors to train the neural network model.
500 realizations are repeated for each setting in order to calculate the coverage probabilities (CP) and the average lengths (AL) of 95\% confidence intervals.
Corresponding numerical results are illustrated in Table \ref{table7}.

\begin{table}[htbp]
	\centering
	\renewcommand{\tablename}{Table}
	\caption{{\small The coverage probabilities (CP) and the average lengths (AL) of
		95\% CIs for pGMC under Example B1 and Example B2.}}
	\begin{tabular}{cccccccccccc}
		\toprule
&            & \multicolumn{4}{c}{Example B1} & & \multicolumn{4}{c}{Example B2}\\
\cmidrule{3-11} 
		&            & \multicolumn{2}{c}{$p=100$} & \multicolumn{2}{c}{$p=200$} & &\multicolumn{2}{c}{$p=100$} & \multicolumn{2}{c}{$p=200$}\\
		\cmidrule{3-6} \cmidrule{8-11}  
        $n_1$    & $n_2$     & CP    & AL    & CP    & AL  & & CP    & AL    & CP    & AL \\
		\midrule
		100   & 100  & 0.874  & 0.135 & 0.860  & 0.133    &   & 0.804  & 0.139  & 0.840  & 0.140 \\
		200   & 200  & 0.918  & 0.093 & 0.882  & 0.092    &   & 0.870  & 0.095  & 0.864  & 0.096  \\
		300   & 300  & 0.926  & 0.074 & 0.912  & 0.074    &   & 0.894  & 0.077  & 0.894  & 0.076  \\
		400   & 400  & 0.930  & 0.064 & 0.926  & 0.064    &   & 0.918  & 0.066  & 0.916  & 0.065  \\
		500   & 500  & 0.946  & 0.056 & 0.942  & 0.057    &   & 0.942  & 0.058  & 0.938  & 0.058  \\
        \bottomrule
	\end{tabular}%
	\label{table7}%
\end{table}%

The results in Table \ref{table7} provide strong evidence that corroborates the asymptotic theory. As the sample size $N$ increases, the empirical coverage probabilities tend to the nominal level 95\% and the average lengths of confidence interval become smaller.

{\section{Real data example}

\subsection{Significant parts on  facial images}

The significance of automatic age prediction of facial images has gained substantial relevance in numerous applications, particularly due to   the rapid growth of social media; see \cite{rothe2015dex}, \cite{antipov2017effective} and \cite{fang2019muti} for instance.
In this subsection, we investigate the significance of three keypoints
on  facial photos for human age prediction.
The data set   is available from   {\emph{www.kaggle.com/datasets/mariafrenti/age-prediction}}  and contains thousands of images, with each image being a three-channel  $128\times 128$ pixel  facial photograph. 
It also records the age of the person  corresponding to each image. Our goal is to  make statistical inferences of the significant predictive features on the images for predicting the age.
\begin{figure}
    \centering
\includegraphics[width=16cm]{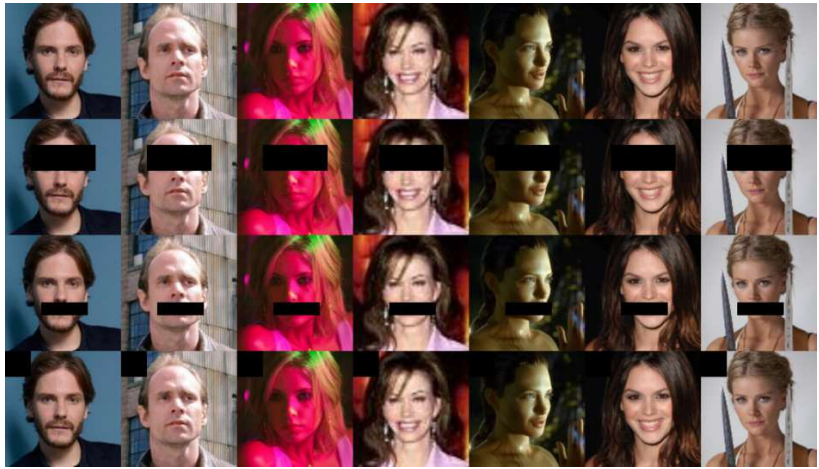}
    \caption{Seven randomly selected original images (the first row). Different parts   are covered in three different cases: case 1 - the 2nd row: eyes region (29:57, 29:100), case 2 - the 3rd row: mouth region (74:89, 40:91), case 3 - the 4th row: top-left corner edge region (0:28,0:28).}
    \label{fig1}
\end{figure}

We randomly select $1000$ images, and artificially cover different key parts on the images (case 1: eyes, case 2: mouths, case 3: top-left corner edge region),  see Figure \ref{fig1} as an illustration. In each covered image,  the covered parts can be considered as the variables of interest (denoted as $W$), while the uncovered parts are denoted as $Z$. We employ a ResNet-18 neural network model and add two fully connected layers at the end of the original network for training. Other hyperparameters keep the same as in the simulation experiments.
 The proposed  procedures along with the other existing methods 
are conducted and the resulting p-values for each case are presented in Table \ref{age_prediction_result}.
\begin{table}[htbp]
  \centering
  \caption{\small{P-values for 
the facial images dataset.}}
    \begin{tabular}{cccccccccc}
    \toprule
          & pMIT  & $\rm pMIT_M$ & pMITe & $\rm pMITe_M$ & pcm   & $\rm pcm_M$ & vim   & DSP   & $\rm DSP_M$ \\
\cmidrule{2-10}    Case 1 & 0.043  & 0.045  & 0.036  & 0.037  & 0.441  & 0.386  & 0.522  & 0.209  & 0.522  \\
    Case 2 & 0.046  & 0.029  & 0.036  & 0.024  & 0.196  & 0.426  & 0.627  & 0.050  & 0.064  \\
    Case 3 & 0.178  & 0.195  & 0.164  & 0.182  & 0.142  & 0.409  & 0.470  & 0.616  & 0.801  \\
    \bottomrule
    \end{tabular}%
  \label{age_prediction_result}%
\end{table}%

While most other testing methods fail to detect the  two significant regions in Cases 1 and 2,
the results in Table \ref{age_prediction_result} show that our pMIT tests consistently reject $H_0$ for the two cases,
suggesting that both eyes  and mouths are significant regions for age prediction in computer vision, which are further visually illustrated  in Figure \ref{fig1}. The corresponding $95\%$ confidence intervals of pGMC for the two regions are $(0.036,0.071)$ and $(0.034,0.055)$, respectively. The region 3 (the top-left corner edge region)  is recognized as not important by all testing procedures, which is consistent with our observation. }

\subsection{Significant gene expression levels}
The second data set is about gene expression and it was once studied in \cite{2006Regulation} and  \cite{2008Adaptive}. It contains 120 twelve-week old male rats from which over 31042 different probes  from eye tissue were measured.  In \cite{2006Homozygosity}, the gene TRIM32 was believed to cause Bandet-Biedl syndrome which is a genetically heterogeneous disease of multiple organ systems including the retina.
\cite{2006Regulation} excluded the probes which were not expressed in the eye or which lacked sufficient variation. As a result, a total of $18976$ probes were considered  as sufficient variables. Among these $18976$ probes, \cite{2008Adaptive} further selected 3000 probes with large variances. We are interested in the genes whose expression would have significant effects on that of gene TRIM32.

We first apply our proposed conditional mean independence test to check whether the 3000 selected probes are  significantly predictive for the expression of TRIM32.  The p-values of the  proposed conditional mean tests are smaller than $0.001$,  
providing a strong evidence that the probes have significant effects on the prediction of the gene expression of TRIM32. 

Based on the results in \cite{2008Adaptive}, there are in total of $19$ significant probes selected by adaptive group LASSO. We care about if any significant probe were omitted or not. Hence, we apply the newly proposed significance test procedures to test whether the remaining $2981$ probes are conditional correlated with the gene expression of TRIM32. We standardize the data and adopt XGBoost and DNN with   three hidden layers  to estimate the condition means. Other existing procedures are also implemented for comparison.  From Table \ref{table10}, it is observed that all the procedures do not reject the null hypothesis at the significance level of $\alpha=0.05$. It means given the $19$ selected probes, other probes do not bring significant additional information to predict the gene expression of TRIM32. Based on the inference above, we conclude that the 19 probes selected by adaptive group LASSO in \cite{2008Adaptive} provide useful information for further biological scientific study in TRIM32.
\begin{table}[htbp]
  \centering
  \renewcommand{\tablename}{Table}
  \caption{The results of partial mean independence tests  for the gene expression data}
  \resizebox{\textwidth}{!}{
    \begin{tabular}{cccccccccccccc}
    \toprule
          & \multicolumn{2}{c}{pMIT} & \multicolumn{2}{c}{$\rm pMIT_M$} & \multicolumn{2}{c}{pMITe} & \multicolumn{2}{c}{$\rm pMITe_M$} & pcm   & $\rm pcm_M$ & vim   & DSP   & $\rm DSP_M$ \\
          & XGB & DNN   & XGB & DNN   & XGB & DNN   & XGB & DNN   &       &       &       &       &  \\
\cmidrule{2-14}    p-values & 0.931  & 0.718  & 0.484  & 0.245  & 0.880  & 0.683  & 0.438  & 0.181  & 0.945  & 0.350  & 0.583  & 0.544  & 0.538  \\
    \bottomrule
    \end{tabular}%
    }
  \label{table10}%
\end{table}%


\section{Conclusion}
In this paper, we propose a new significance test for the partial mean independence problem  based on machine learning methods and data splitting, which is applicable for high dimensional data.  The pMIT test statistic converges to the standard chi-squared distribution under the null hypothesis while it converges to a normal distribution under the fixed alternative hypothesis.  Power enhancement and algorithm stability based on multiple data splitting are also discussed.
When the partial mean independence test is rejected, we propose
a new partial dependence measure, called partial Generalized Measure of Correlation (pGMC), based on the decomposition formula of the conditional variance.
We derive its theoretical properties and construct confidence interval.  
As an important special case when there is no conditioning random object, we also 
investigate significance testing of potentially high dimensional covariates $X$ for the conditional mean of $Y$ in a model-free framework.  In the future, it is interesting to extend the proposed ideas to study conditional/partial quantile independence, conditional/partial independence problems.  

\noindent
\textbf{Supplementary materials.} Additional simulation results including the finite sample performance of the conditional mean independence
test, and all theoretical proofs  are included in the supplementary materials.

				{ \baselineskip 14pt\bibliographystyle{asa}
					\bibliographystyle{asa}
					\bibliography{bibliography}}
				
			\end{document}